# Potential of electrostatic micro wind turbines

M. Perez[1,2,4], S. Boisseau[1,2], N. Perez[3], and J. L. Reboud[1,4]

[1] Univ. Grenoble Alpes, F-38000 Grenoble, France
[2] CEA, Leti, Minatec Campus, 17 rue des Martyrs – F-38054 Grenoble, France
[3] ENS-Paris, 45 rue d'Ulm – F-75005 Paris, France
[4] Univ. Grenoble Alpes, G2Elab, F-38000 Grenoble, France

E-mail: matthiasperez1@gmail.com

**Abstract.** This paper presents the physical operating principles of several micro wind turbines based on different aerodynamic forces: drag-type Vertical Axis Wind Turbine (VAWT) and lift-type Horizontal Axis Wind Turbine (HAWT). All these devices share the similarity of exploiting the same mechanical-to-electrical conversion: the electrostatic conversion. This type of conversion is based on capacitance variations induced by the motion between a rotor and a stator and requires a source of polarization. We will focus our study on two technologies to polarize the capacitive structure: the use of electrets and the exploitation of triboelectricity. Some experiments conducted in a low-speed wind tunnel between 0 and 20m.s$^{-1}$ have highlighted power flux densities from 0 to 150μW.cm$^{-2}$ corresponding to power coefficients of 0 and 9% respectively. Among these results, we can especially retain an ultra-low speed operation, which has never been reached until now, in terms of speed and efficiency (9% of efficiency at 1m.s$^{-1}$). Finally, we will end up comparing different types of circuits to supply a temperature/acceleration sensor, in order to complete the energy harvesting chain.

**Introduction, state of the art**

Fluid flows concern a large variety of common situations and daily phenomena, for instance: winds, ventilation and filtration systems flows in buildings (<4m.s$^{-1}$), or simply the perceived stream when we walk (<2m.s$^{-1}$), run (<5m.s$^{-1}$), cycle (<20m.s$^{-1}$), and even when we breathe (<1.5m.s$^{-1}$). As the kinetic power of an airflow is given by equation (1), with ρ≈1.225kg.m$^{-3}$ the air density, S the collecting surface and U the airflow velocity.

$$P_{kin} = \frac{1}{2}\rho S U^3 \quad (1)$$

The power flux density of kinetic energy related to the airflows described above range from 0 to 900mW.cm$^{-2}$, which is in perfect agreement with the energy required by a lot of electronics applications and especially communicating sensors (Table 1).

| Object to supply | Power associated | References |
|---|---|---|
| Microcontroller chip in standby mode | 200nW | PIC18F1220/1320 |
| Quartz watch | 700nW | Seiko AGS |
| Temperature sensor | 6μW at 1Hz | ST HTS221 |
| Pressure sensor | 12μW at 1Hz | MS5541C |
| Temperature measurement and BLE transmission | 110μW at 1Hz | [1] |
| Recharge of a AA battery | 300μW during 1 year | Duracell UltraPower AA 2500mAh |
| Light emitting diode (LED) | 80mW | Vishay Ø3mm TTLE4401 |

**Table 1. Some examples applications needing electric energy.**

Unfortunately, all this kinetic power cannot be extracted entirety. In fact, we can prove that we can extract, in the best case, 16/27 of the kinetic power, it is the so-called "Betz limit" [2]:

$$\eta_{Betz} = \frac{16}{27} \quad (2)$$

For more than a decade, a lot of researchers have tried to develop efficient miniature systems. Many harvesters have been developed, with a common thread to exploit the motion of a solid element. They can be separated into two main categories:
- Rotational harvesters using the kinetic energy from the airflow to create a rotary motion of a rigid rotor.
- Aeroelastic harvesters using the kinetic energy from the airflow to induce a vibratory motion of a flexible element.

All these prototypes have been coupled to a lot of electromagnetic ([3][4][5][6][7][8][9][10][11][12][13][14][15][16][17]) and piezoelectric ([18][19][20][21]) converters but only a few electrostatics ([22][23][24][25]). Some authors have been able to develop high speed devices that could be used in the automotive sector or in aeronautics. However, it appears that most authors have tried to develop low speed devices for applications previously cited. Overall, one notes that the miniaturization leads to a reduction of the efficiency, with a limit of 10% for cm-scale devices at speeds exceeding $2m.s^{-1}$.

The purpose of this paper is to assess the potential of the electrostatic conversion with two types of cm-scale rotating wind turbines (drag-type VAWT and lift-type HAWT) with diameters from 2.5cm ($5cm^2$) to 4cm ($12.6cm^2$). In addition, we will focus our study on low speed airflows (<$20m.s^{-1}$) and electrostatic conversion with two types of polarizable elements (electrets and triboelectrets). Finally, we will end up comparing different types of circuits to supply a temperature/acceleration/magnetic field sensor.

## 1. Aeromechanical conversion at ultra-low Reynolds number

This section gives an overview of various ways to convert the kinetic energy of an airflow into usable mechanical energy. Aeromechanical converters can be divided into different categories according to the force used to produce the mechanical energy and the nature of the mechanical motion. In this work, we will focus on drag-type VAWT and lift-type HAWT.

Drag-type wind turbines use the drag force $F_D$ (more specifically the form drag) to induce a mechanical rotation. Figure 1a presents a Savonius wind turbine which has the strong advantage to be very simple to design and manufacture, thus allowing to reduce the associated expenses. As the wind turbine is rotating, the relative speed between the rotating blade and the flow depends on the distance from the hub (r). Their main drawback is the serious restriction of their rotational speed ($\lambda=\omega R/U \leq 1.5$). Given the relatively small drag coefficients[1] $C_D$ (at best, $C_D \approx 1.4$ for a half hollow cylinder [26]), the drive torque will also be low, limiting the efficiency of these wind turbines to approximately 15% (without taking of mechanical losses into account). It is also interesting to note that if the direction of the airflow is well known, a carter can also be added to optimize the driving drag on the second quarter of the trajectory and delete the resistive drag on the last quarter of the trajectory (when the blade moves in the direction opposite to the airflow), thus slightly increasing the overall efficiency (Figure 1a).

The lift-type wind turbines use the lift force $F_L$ to induce mechanical rotation (Figure 1b). They are the most commonly used on a large scale (large wind farms) as well as on a small scale (state of the art presented in the introduction). A lift-type wind turbine consists in a central hub on which one or several blades are attached (Figure 1c). For obvious reasons of cylindrical periodicity, the different parameters will therefore depend on the distance r from the rotational axis and the mechanical design of an HAWT will therefore require the knowledge of 5 parameters:
- The radius of the windmill R
- The number of blades $N_p$
- The evolution of the chord $c(r)$
- The evolution of the blade angle $\beta(r)$
- An aerodynamic profile (ex: flat plate, thin camber plate, NACA0012...)

---

[1] $C_D$ is the drag coefficient, dimensionless version of the drag force.

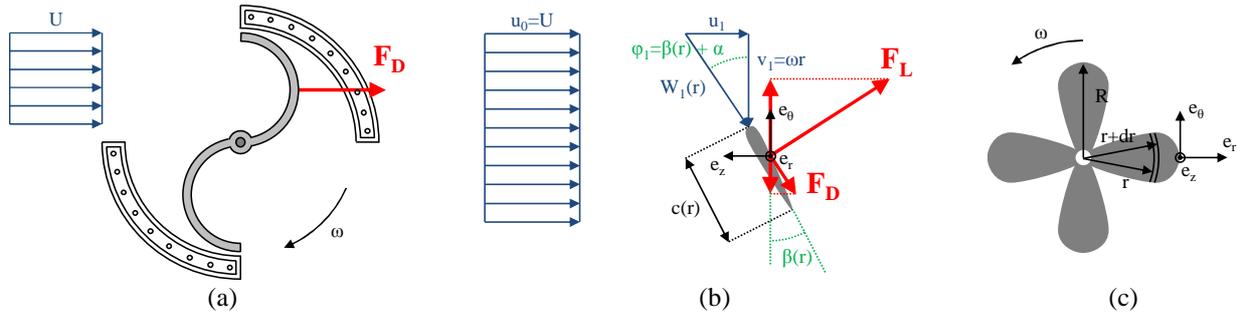

**Figure 1: Working principle of two different wind turbines. (a) Top view of a drag-type VAWT Savonius turbine and its carter. (b) Lateral view and (c) front view of a lift type HAWT.**

### 1.1 Lift and drag at ultra-low Reynolds number

As mentioned in the introduction, the cumulative constraints of size (<4cm) and speed (<20m.s$^{-1}$) impose ultra-low Reynolds numbers (Re≈UR/ν<10$^4$). This operating area directly implies an increase of viscous drag and a decrease of form drag. Figure 2a shows the evolution of the total drag coefficient on an infinite cylinder and we can clearly see that the diminution of the Reynolds number leads to an increase of the viscous drag coefficient (red arrow in Figure 2a). On the contrary, the form drag coefficient (green arrow in Figure 2a) tends to decrease step by step: first when the turbulent boundary layer with detachment (Re>10$^3$) is replaced by a laminar boundary layer with detachment (10$^1$<Re<10$^3$) which is then itself replaced by a laminar boundary layer without detachment (Re<10$^1$).

Figure 2b presents the progression of the lift force on an aerodynamic profile (NACA0012) for different angles of attack (α). We can notice that as the Reynolds number decreases, the angle of attack which maximizes the lift force declines as well as the maximum lift force itself. This evolution is due to the fact that the boundary layer on the upper surface of the profile becomes fragile with the diminution of the Reynolds number and a detachment appears at a lower angle of attack. Since the lift force is theoretically proportional to the sinus of the angle of attack, which is experimentally true before the detachment of the boundary layer (Figure 2b), the maximum lift decreases accordingly.

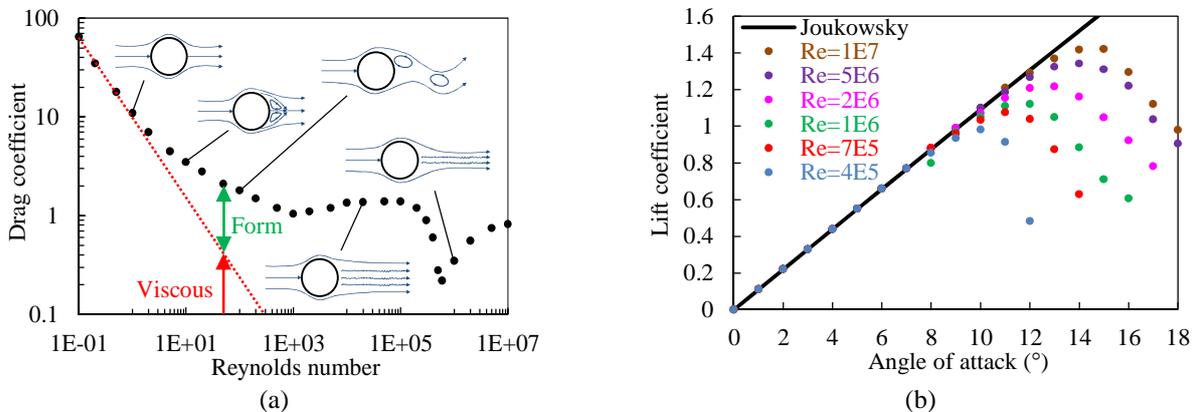

**Figure 2: (a) Effect of the Reynolds number on drag forces on an infinite cylinder (viscous + form) [26]. (b) Lift force on an NACA0012 profile as a function of the attack angle for various Reynolds numbers [27].**

### 1.2 Design of lift based micro wind turbines
#### 1.2.1 Ideal design according to Betz

Let's assume that the wind turbine does not induce any vortex wake and that the drag of the blades is negligible. Figure 1b represents the velocity triangle in the rotary coordinate system at the rotor level. The local wind speed actually perceived by the blade ($w_1$) is therefore equal to the sum of the wind speed at the level of the rotor (ideally and according to Betz: $u_1=2U/3$ [2]) and the speed due to its own rotation ($v_1=\omega r$). By introducing the tip speed ratio $\lambda=\omega R/U$, we can write:

$$\begin{cases} u_1 &= \dfrac{2}{3}U \\ v_1 &= \dfrac{\lambda r}{R}U \\ w_1 &= \left(\sqrt{\dfrac{4}{9} + \dfrac{\lambda^2 r^2}{R^2}}\right)U \end{cases} \qquad (3)$$

In order to get an optimal angle of attack $\alpha_{opt}$ seen by the blade (generally the angle which maximize the lift-to-drag ratio $C_L/C_D$ and not the angle which maximize the lift), the evolution of the blade angle $\beta(r)$ must verify the following equation:

$$\beta(r) = \varphi_1 - \alpha_{opt} = \operatorname{atan}\left(\frac{2R}{3\lambda r}\right) - \alpha_{opt} \qquad (4)$$

Considering a blade element with a thickness dr located at a distance r from the rotor axis, the elemental lift force $dF_L$ projected onto the axis $e_\theta$ (Figure 1c) is:

$$dF_\theta(r) = \frac{1}{2}\rho w_1^2 c dr C_L \sin(\varphi_1) \qquad (5)$$

The elemental power converted by this ring section dr is consequently equal to:

$$dP(r) = \cdots = \frac{1}{3} N_p \rho \left(\sqrt{\frac{4}{9} + \frac{\lambda^2 r^2}{R^2}}\right) \frac{\lambda r}{R} U^3 c C_L dr \qquad (6)$$

But, according to Betz, the maximum power produced by this ring section is equal to:

$$dP(r) = \frac{16}{27} \times \frac{1}{2}\rho U^3 \times 2\pi r dr \qquad (7)$$

It can be de deduced the evolution of the optimal chord as a function of the radius r:

$$c(r) = \frac{16\pi R}{N_p C_L} \times \frac{1}{9\lambda \sqrt{\dfrac{4}{9} + \dfrac{\lambda^2 r^2}{R^2}}} \qquad (8)$$

### 1.2.2 Wake losses

Betz's model does not take into account the vortex wake that appears downstream of the wind turbine with the deflection of the airflow by the rotor blades. To be very precise, this deflection is perpendicular to the relative speed $w_1$ and is essential in order to obtain a lift force. Considering these new conditions and after some calculations detailed in [28], we can deduce the new evolution of the blade angle recommended by Glauert:

$$\beta(r) = \frac{2}{3}\operatorname{atan}\left(\frac{R}{\lambda r}\right) - \alpha_{opt} \qquad (9)$$

And the corresponding evolution of the optimal chord:

$$c(r) = \frac{16\pi r}{N_p C_L}\sin^2\left(\frac{\operatorname{atan}\left(\frac{R}{\lambda r}\right)}{3}\right) \qquad (10)$$

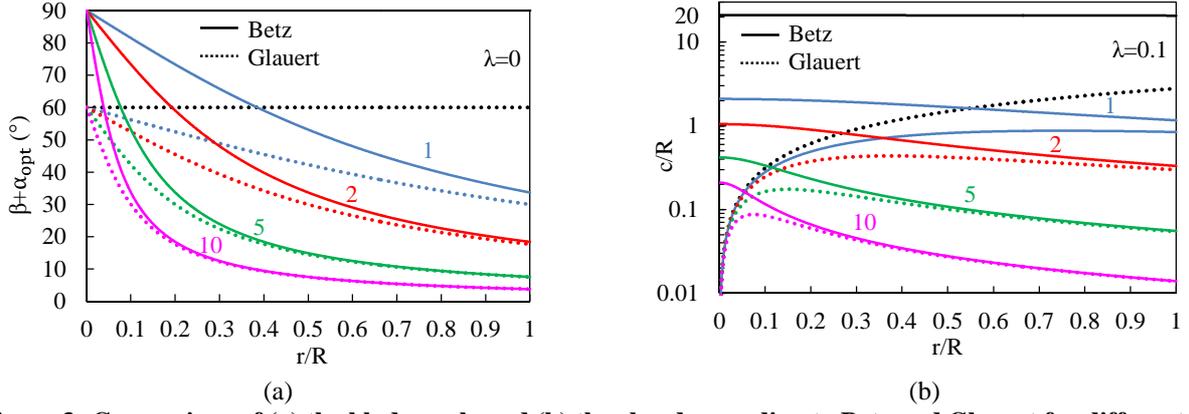

(a)      (b)

**Figure 3: Comparison of (a) the blade angle and (b) the chord according to Betz and Glauert for different tip speed ratio $\lambda$ ($N_P$=4, $C_L(\alpha_{opt})$=1).**

Finally, by integrating the elemental power on the total area of the wind turbine, we can deduce the wake efficiency:

$$\eta_{wake} = \frac{27\lambda}{4} \int_0^1 \frac{\left(\frac{r}{R}\right)^2 \sin^3\left(\frac{2}{3}\text{atan}\left(\frac{R}{\lambda r}\right)\right)}{\sin^2\left(\text{atan}\left(\frac{R}{\lambda r}\right)\right)} d\left(\frac{r}{R}\right) \tag{11}$$

This integral can only be calculated numerically. Figure 4 presents the results obtained (blue curve). This representation of the evolution of the wake losses as a function of the tip speed ratio allow us to realized that it is preferable to operate at high tip speed ratio in order to minimize the wake losses.

### 1.2.3 Drag losses

In the previous calculations, the drag force was neglected, we are now going to consider its adverse effect through the intermediary of a torque loss. I order to take it into account, we just have to replace the element $C_L\sin(\varphi_1)$ by $C_L\sin(\varphi_1)-C_D\cos(\varphi_1)$ in the previous calculations, Which is equivalent to introduce à local profile efficiency:

$$\eta_{drag}(r) = \frac{C_L \sin(\varphi_1) - C_D \cos(\varphi_1)}{C_L \sin(\varphi_1)} = 1 - \frac{3C_D \lambda r}{2C_L R} \tag{12}$$

By integrating this local efficiency on the total area of the wind turbine, the overall profile efficiency can be deduced:

$$\eta_{drag} = \frac{1}{\pi R^2} \int_0^R \eta_{drag}(r) \times 2\pi r dr = 1 - \frac{C_D \lambda}{C_L} \tag{13}$$

We can conclude that the tip speed ratio $\lambda$ cannot exceed the value of $C_L/C_D$ (lift-to-drag ratio or glide ratio) and that there is a maximum of power ranging between 0 (high torque and low rotational speed) and $C_L/C_D$ (low torque and high rotational speed).

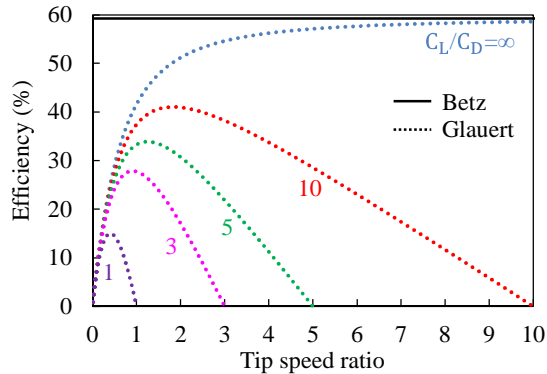

**Figure 4: Global efficiency considering the wake (blue curve) and drag losses.**

If we reduce the size of the wind turbine, we have previously seen that the viscous drag will become more and more important (sections 1.1), the lift-to-drag ratio of the blades will inexorably fall, just as the performances of the wind turbine.

### 1.2.4 Tip losses

This loss is due to the fact that the length of the blades is not infinite and that the pressure difference between the upper and the lower side of the blades cannot be maintained at the tip of the blades. It is particularly notable at low tip speed ratios and with a small number of blades, which implies low aspect ratios. This way we can define the following tip efficiency [29]:

$$\eta_{tip} = \left(1 - \frac{0.92}{N_p \sqrt{\lambda^2 + \frac{4}{9}}}\right)^2 \tag{14}$$

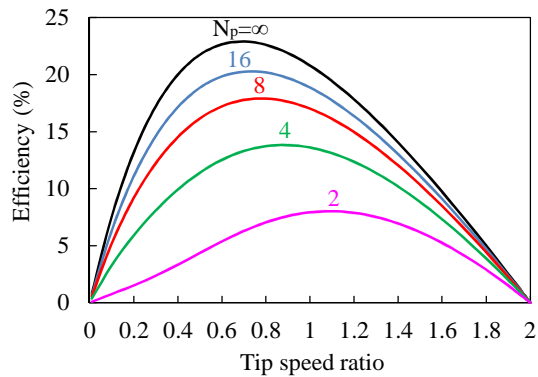

**Figure 5: Global efficiency, taking into account the tip losses ($C_L/C_D=2$).**

### 1.2.5 Bearing losses

This last loss is common to all types of wind turbines because it is due to the friction inside the bearings that we used. The corresponding torque can be decomposed into three contributions:

- The friction torque of Coulomb $C_{Coul}$ which has itself two origins: a static friction which tends to keep a body in a static state and a dynamic friction which tends to slow down a body in motion. This torque is proportional to the actual contact area (strong link with the roughness) and is strongly influenced by the couple of materials used.
- The viscous friction torque appears by friction between the bearing and the air present all around. Given the size of our bearings, the flow around the bearing can be considered laminar, hence the friction torque proportional to the rotational speed: $C_{visc}\omega$. However, this contribution is negligible in comparison to the previous drag losses.
- The viscous friction torque of Stribeck comes up when we put in touch two lubricated solid. It therefore seems that the resistive torque associated decreases exponentially with the rotational speed. Nevertheless, this contribution should only be regarded for lubricated bearings, which is not the case in this work.

$$C_{roul}(\omega) = C_{coul} + C_{Stri} \exp\left(-\frac{\omega}{\omega}\right) + C_{visc}\omega \approx C_{coul} \tag{15}$$

The power dissipated by the bearing takes the following form:

$$P_{bear} \approx C_{coul}\omega \tag{16}$$

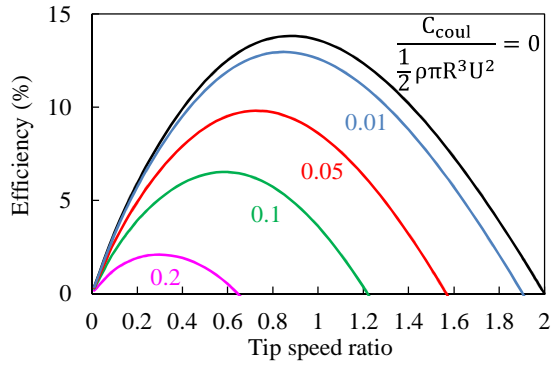

**Figure 6: Global efficiency considering the bearing losses ($C_L/C_D=2$, $N_p=4$, $C_{stri}=C_{visc}=0$).**

In our effort to miniaturize the devices (R is decreasing), it is thus necessary to maximize the rotational speed ω in order to keep an optimal tip speed ratio $\lambda=\omega R/U$. As a consequence, the bearing losses will be strongly accentuated.

### 1.2.6 Choice of the blade profile

The selection of the blade profile is very important because it determines the lift and drag coefficients as well as the optimal angle of attack, which will dictate the design of the wind turbine (chord, blade angle, rotational speed and efficiency). The objective is therefore very simple: choose a profile with the highest lift-to-drag ratio $C_L/C_D$ in our operating area ($Re<10^4$).

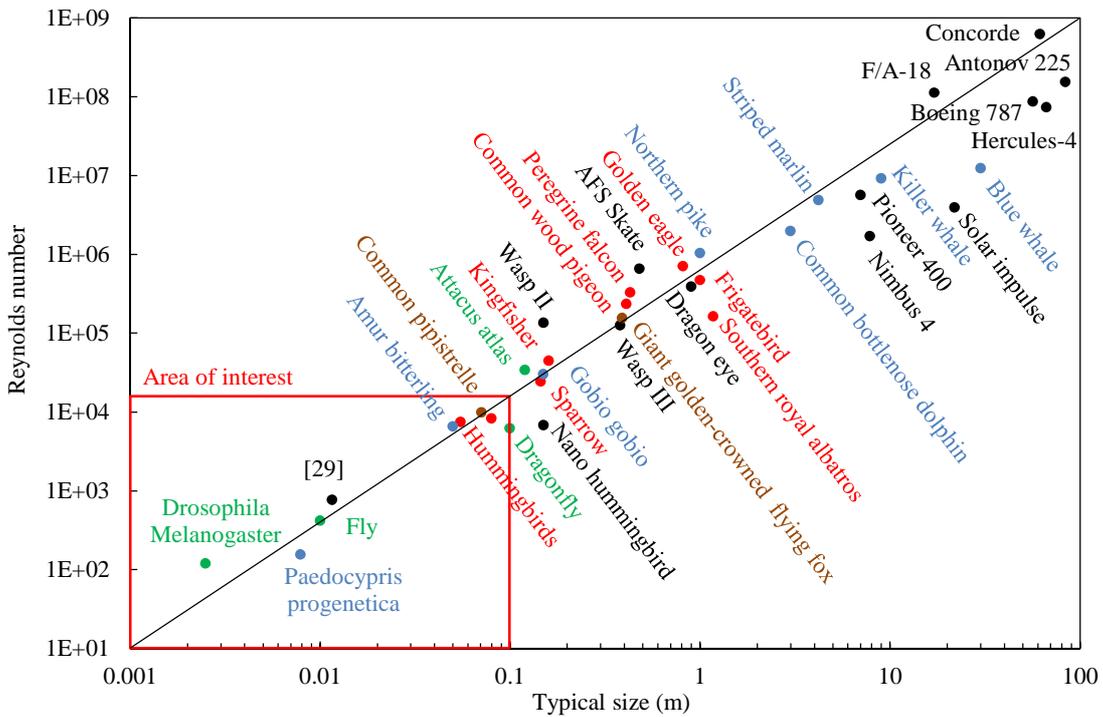

**Figure 7: A few examples of animals and existing structures with the Reynolds number associated. We can find marine animals in blue, some birds in red, insects in green, terrestrial mammals in brown and human structures in black.**

One of the first ideas was to look at Reynolds number aerodynamic profiles ($Re>10^6$) which are used in aviation and large scale wind power (Figure 8h and Figure 8i). In this operation area, millions of tests have been perform for more than a century, which contributed to obtain some profiles with a lift-to-drag ratio higher than 100. These profiles are commonly thick, smooth, more or less cambered with a rounded leading edge and a sharp trailing edge. On its side, nature has taken several millions of years to develop reliable profiles that allow the displacement of a lot of birds, mammals, and cartilaginous fishes at high Reynolds number ($10^4<Re<10^7$ in Figure 7). In most cases, the leading edge of their wings or fins is also thick because it contains bones or cartilage, muscles, tendons as well as ligaments. Then, the profile becomes finer to end on a sharp trailing edge thanks to thin layers of feathers or skin (Figure 8d and Figure 8f).

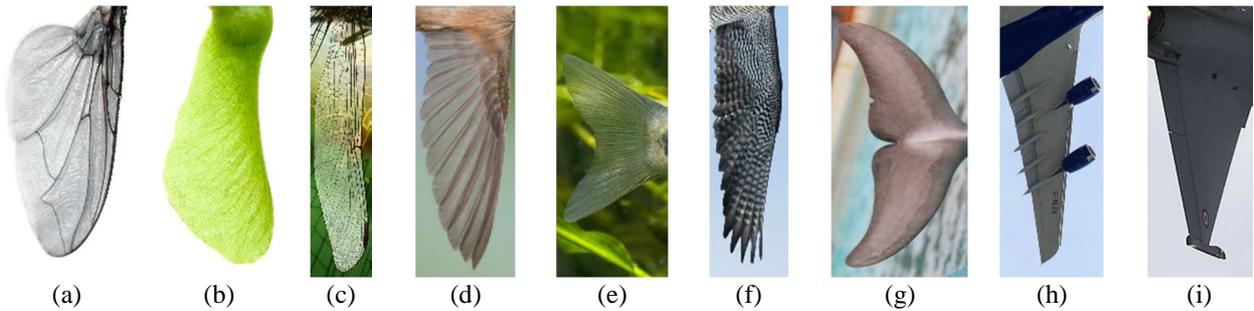

| (a) | (b) | (c) | (d) | (e) | (f) | (g) | (h) | (i) |

**Figure 8: A few examples of aerodynamic profiles by increasing order of Reynolds number: (a) fly's wing (Re≈4×10$^2$), (b) maple seed (Re≈1×10$^3$), (c) dragonfly's wing (Re≈6×10$^3$), (d) hummingbird's wing (Re≈7×10$^3$), (e) gobio gobio's fin (Re≈3×10$^4$), (f) falcon's wing (Re≈3×10$^5$), (g) dolphin's fin (Re≈2×10$^6$), (h) airliner's wing (Re≈9×10$^7$), (i) supersonic aircraft's wing (Re≈6×10$^8$).**

However, the constraints of our study impose much lower Reynolds numbers (Re<10$^4$). In this specific operation field, the viscous drag is strong (section 1.1), the lift force is potentially limited by a weak laminar boundary layer (section 1.1); which leads to an overall reduction of the lift-to-drag ratio of the aerodynamic profiles (Figure 9). In addition, investigation at low Reynolds number is far less advanced and optimal profiles at high Reynolds number are not effective at low Reynolds numbers (Figure 9). In fact, in this operation zone, it appears that thin profiles ([31], [32]), rough profiles [33] and slightly cambered profiles (Figure 9) offer better performances. In nature, such Reynolds numbers are associated to insect's flight and bony fish swimming (Figure 7). Indeed these animals possess very thin wings or fins from the leading edge to the trailing edge, while being stiffened by a network of ribs/veins or bone spurs (Figure 8a, Figure 8c and Figure 8e). It is therefore towards this type of profiles that we will orient the design of our micro wind turbines.

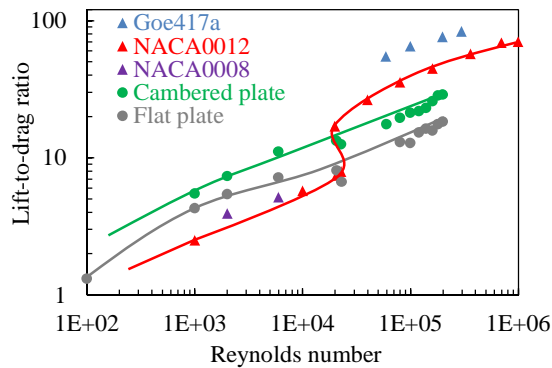

**Figure 9: Evolution of the lift-to-drag ratio for different profiles with the chord Reynolds number ([27], [34], [35], [36], [37], [38] and [39]).**

### 1.2.7 Overview

Our current knowledge enables us to have a global view on the design of lift-type micro wind turbines:
- The radius R is set to a few centimeters by our constraints of size, which should be sufficient to produce a few hundreds of microwatts at low speed (1-3m.s$^{-1}$). However, these choices require operating at ultra-low Reynolds number (between 600 and 6000).
- Such Reynolds numbers invite us to select thin blade profiles (thickness/chord ratio in the order of a few percent) and slightly cambered (by a few percent too). The structure should nevertheless be robust enough to withstand extreme conditions of input speed (10-20m.s$^{-1}$) and rotational speed (~5000rpm). According to the state of the art, such profiles should provide lift-to-drag ratios in the order of a few units (1<$C_L/C_D$<7) for Reynolds number lower than 10$^4$. The optimal angles of attack $\alpha_{opt}$ fluctuate around 5° according to the aspect ratio, with lift coefficients associated ranging between 0.2 ($\alpha$=2°) and 1 ($\alpha$=10°) [39].
- The number of blades $N_p$ is also an important parameter subject to compromise. In fact, a small number of blades leads to high chord lengths, which causes a favorable increase of the chord Reynolds number and the lift-to-drag ratio associated but also an unfavorable decrease of the aspect ratio. On the contrary, a high number of blades leads to small chord lengths, which causes an unfavorable decrease of the chord Reynolds number and the lift-to-drag ratio associated but limits the tip losses.
- The evolution of the chord c(r) is determined by $\alpha_{opt}$, $C_L$ and $N_p$. As we just mentioned, the low tip speed ratios force us to compensate by increasing the motor torque. The chords of the blades need to be increased wherever possible, so that their length will be the same order of magnitude as the size of the wind turbine, i.e. $N_p \times c(r) \sim 2\pi r$. This last practical limit may bring us to favor higher angles of attack ($\alpha_{opt} \approx 10°$) allowing higher lift forces ($C_L \approx 1$), even if it leads to a reduction of the lift-to-drag ratio...
- The evolution of the blade angle $\beta(r)$ can thus be deduced from the knowledge of $\lambda$ and $\alpha_{opt}$.

Finally, the mechanical power available (for our electrostatic converter) is strongly limited by all the losses that we just mentioned, which tend to increase with the constraints of size and speed of our study. This explains the poor yields in the state of the art (<10%).

## 2. Electrostatic conversion as a mechanical-to-electrical conversion

As the electromagnetic conversion is based on a variation of magnetic flux through an inductive circuit, the electrostatic conversion relies on a variation of electric field in a capacitive circuit. In order to reach this target, it is necessary to use at least two electrodes in relative motion, which lead to capacitance variations and a source of polarization. Globally, the electrostatic conversion has the advantage of not inducing losses compared to electromagnetic converters (winding Joule, hysteresis, eddy current losses…). In this work, all the converters have been achieved according to a "slot effect" configuration (Figure 10a and [40]) where the electric current flows between the fixed electrodes of the device. We will also use two types of polarization: by addition of electrets [23][24] or by triboelectricity [25].

### 2.1 Electret-based electrostatic conversion

One option is to to use an electret-based electrostatic converter that consists in 2N fixed electrodes and N mobile counter-electrodes spaced by a distance e. The relative motion between the electrodes and the counter-electrodes therefore induces capacitance variations required by the electrostatic conversion. Then, a thin electret layer is deposited on all the fixed electrodes (N=1 in Figure 10a). Electrets are simple dielectric layers that we charged by Corona discharge in order to obtain high voltage stable electrets (Figure 10b). As we will see below, electret-based conversion is highly advantageous for a low speed startup compared to the electromagnetic conversion. This advantage derives from the fact that electrets are very thin and lights compared to permanent magnets and no static torque appears. However, we will see that these converters offer a weak coupling (airgap relatively high and polarization voltage limited for the electrets) which restricts its use to low speed applications.

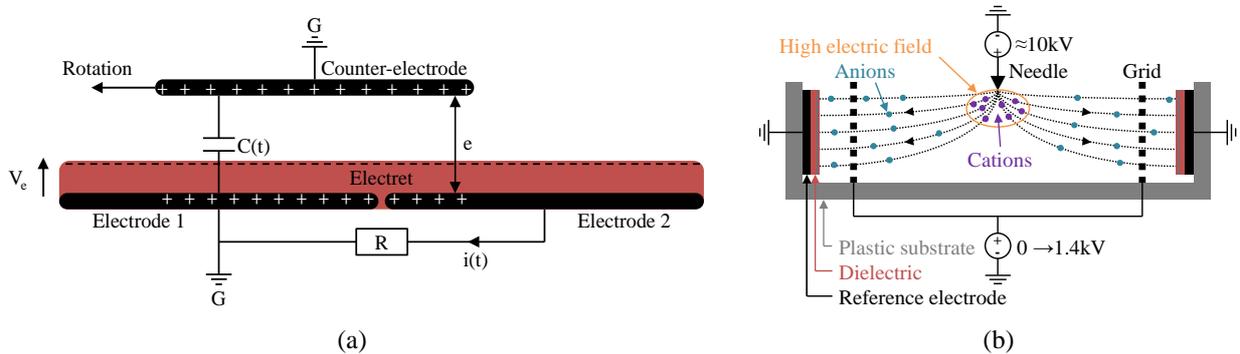

**Figure 10. (a) Electret-based converter with a simple load R: slot effect connection scheme. (b) Cylindrical corona discharge to polarize the electrets of the stator.**

### 2.2 Triboelectret-based electrostatic conversion

The second option is to use a triboelectret-based electrostatic converter previously presented in [25]. Just like before, we will use 2N fixed electrodes but this time $N_m$ thin membranes ($N_m \leq N$) able to come into contact from an electrode to another one (Figure 11). The mechanical contact between these two different materials provokes an exchange of electric charges by triboelectricity. For Teflon FEP electrets that we use in this work, negative electric charges are created on the dielectric surface (<0.25mC.cm$^{-2}$) and the same amount of positive electric charges appears on the electrodes surface. The operation is then similar to an electret-based operation with a circulation of the electric charges between the different electrodes. One of the benefits of this type of device relies in the fact that strong capacitance variations can be obtained without needing an accurate control of the airgap (as is the case of an electret-based operation). As regards polarization, surface potentials obtained are slightly lower compared to what can be achieved with electrets (570V vs 1400V for a 125µm-thick Teflon FEP electret). In addition, this technique provide a self-polarization of the capacitive structure, making these converters much "electrically stronger" than electret-based converters (which are very sensitive to dust, humidity…). By contrast, triboelectret-based conversion intrinsically requires friction to operate. The corresponding friction torques are generally very high, which tends to strongly decrease the startup speed, the operating tip speed ratio, hence the performances.

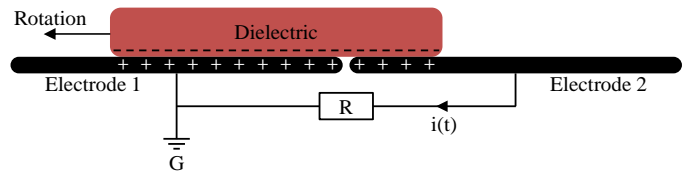

**Figure 11. Triboelectret-based converter with a simple load R.**

## 3. Prototypes and experimental results
### 3.1 Test bench and instrumentation

In order to test our prototypes at very low speed (<1m.s$^{-1}$), we developed a wind tunnel with a constant square cross-section of 36cm×36cm for an overall length of 1.72m. Nine low flow fans (ebm-papst 4414N) of 12cm×12cm were placed downstream from the test section (Figure 12a) in order to operate in suction. Two honeycomb filters including circular alveolus of 4mm in diameter for a depth of 3cm were located downstream and upstream from the test section in order to generate an airflow as laminar as possible (Figure 12c). A convergent was also placed downstream from the wind tunnel to reduce potential turbulences. Finally, the velocity measurement was achieved by a hot wire anemometer (Fisher Scientific AM-4204) with a resolution of 0.1m.s$^{-1}$, located a few centimeters above the prototype (Figure 12b).

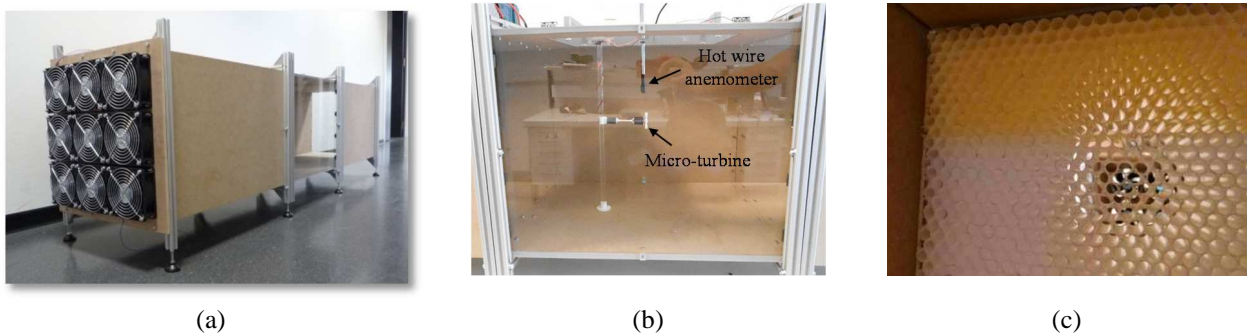

(a) (b) (c)

**Figure 12. (a) Pictures of the wind tunnel: (a) overview, (b) test section and (c) honeycomb filter.**

### 3.2 Overall functioning

In the case of an electret-based drag-type wind turbines, the electrostatic converter takes the form of two discs located below and above the rotor (Figure 13a). Circular sections of electrodes are located on the mobile disc while the fixed part is composed of electrodes covered by a layer of electrically charge dielectric (the electret).

In the case of an electret-based lift-type wind turbine, the basic idea remains the same but this time the electrostatic converter takes the form of concentric cylinders as presented in Figure 13b.

The last devices considered in this work are the triboelectric lift-type wind turbines. The structure includes a rotor of axial wind turbine on which we will incorporate $N_m$ (≤N) thin dielectric membranes, as illustrated in Figure 13c. Each one of these membranes has a length L=πR/N, a width H and a thickness h. Like electret-based lift-type wind turbines, the cylindrical stator includes 2N electrodes and is positioned around the rotor.

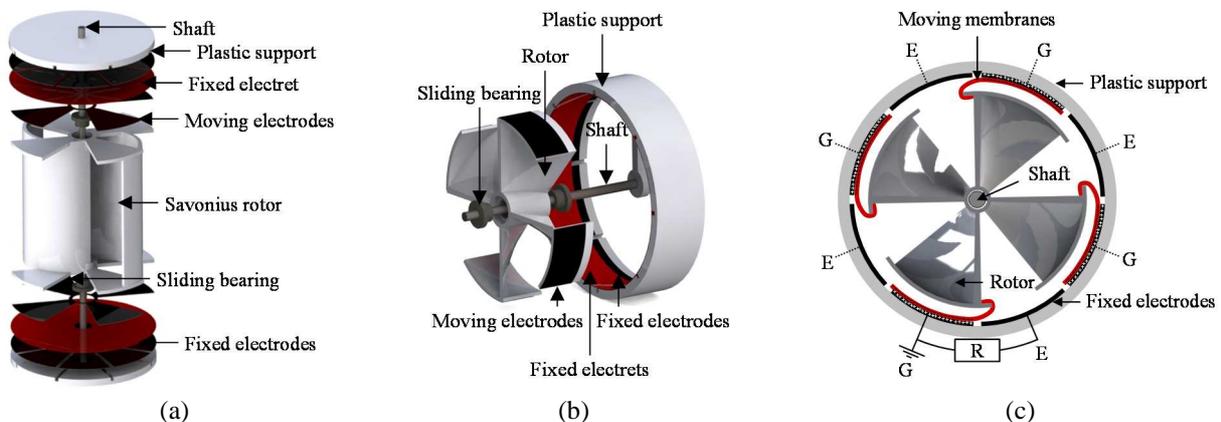

(a) (b) (c)

**Figure 13. (a) Drag-type Savonius wind turbine and circular electret-based electrostatic converter (N=4 moving electrodes on each side). (b) Lift-type wind turbine and cylindrical electret-based electrostatic converter (N=4 moving electrodes). (c) Operation scheme of a lift-type wind turbine and its cylindrical triboelectric converter ($N_m$=4 membranes and 2N=8 electrodes).**

### 3.3 Prototypes fabrication and dimensioning and fabrication

The mechanical base of the different prototypes has been manufactured by stereolithography (Figure 14), with blade thicknesses of 750µm in order to get solid devices. For electret-based micro wind turbines (drag-type and lift-type), the counter electrodes of the stator are made of 60µm-thick copper covered with a 125µm-thick Teflon FEP layer, while the electrodes of the rotor are made of a thin silver layer of a few micrometers. The electret is polarized by an annular corona discharge (Figure 10b) respectively at 600V for R=15mm and 1400V for R=20mm. Note that this is the first time the good operation of the annular corona discharge is validated. For triboelectret-based lift-type micro wind turbines, the electrodes of the stator are also made of 60µm-thick copper while the membranes are made of 50µm-thick Teflon FEP layers.

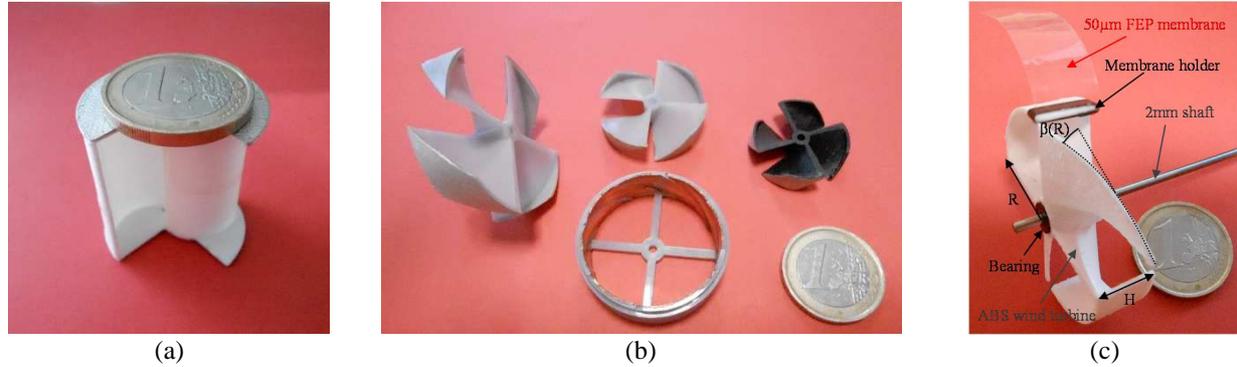

**Figure 14. (a) Picture of a drag-type Savonius wind turbine and its electret-based electrostatic converter (R=15mm, H=30mm, N=2). (b) Picture of 3 electret-based lift-type wind turbines (H=R/2, N=4). (c) Picture of a lift-type triboelectric wind turbine (R=20mm, H=10mm, N=4, $N_m$=1)**

Regarding lift-type micro wind turbines (electret-based or triboelectret-based), minor changes were required compared to the optimal design of Glauert. Regarding the evolution of the blade angle β(r) and the chord c(r), it is important to note that the design proposed in section 1.2 is valid for a given tip speed ratio (related to the rotational speed ω=λU/R). However, considering that the rotor is initially at rest and has to speed up to reach its design tip speed ratio, the initial conditions (λ=0) are therefore different from the design conditions (λ≠0). This difference is further emphasized by the magnitude of the design tip speed ratio. At large scale, this is not a problem because the friction torque of the bearing is very small and the alternator can be decoupled from the windmill during the rotational speed-up (a small motor torque is therefore sufficient to achieve the desired tip speed ratio). In our case, the friction torque of the bearing is far from negligible and the electrostatic converter cannot be decoupled from the micro wind turbine during the rotational speed-up. It results that the motor torque induced by the lift of the blades must be greater than the sum of the resistive torques for tip speed ratio lower than the design tip speed ratio. If this is not the case, the rotor will remain blocked in its fixed position and no energy can be extracted. We therefore decided to use the evolution of the blade angles recommended by Betz (Figure 15a) which allows to achieve the most interesting angles of attack at the hub level when the rotor is stopped (close to 5-10°), and thereby to foster the rotor startup. In addition, the chord length has been slightly modified in the area close to the hub and increased on the outside of the turbine (Figure 15b) in order to increase the motor torque during the startup and to fix the height of the rotor (H=c(r)×sin(β(r))=cte). Moreover, this last characteristic makes our prototypes very suitable for micro-fabrication processes by removing material (laser cutting, plasma, LIGA...) or additives (3D printing).

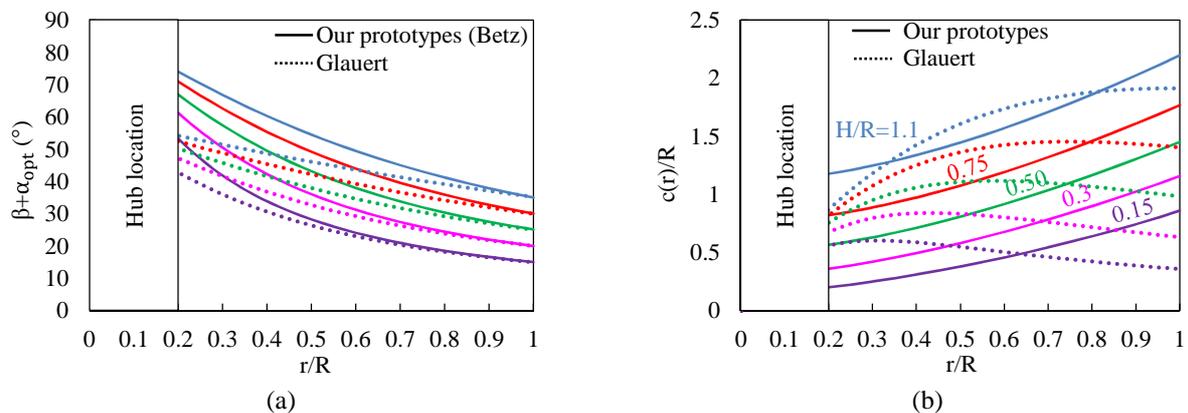

**Figure 15: Parameters of the different prototypes: radial evolution of (a) the blade angles and (b) the chord length ($N_p$=4, $α_{opt}$=5°, $C_L(α_{opt})$=0.6).**

### 3.4 No-load rotational speed

Figure 16a shows the rotational speeds achieved on Savonius wind turbines of 20mm radius and heights within a range from 10mm to 40mm and realized by 3D printing. The higher the height of the turbine, the higher the drag force is and therefore more able to offset the friction torque of the bearing. The rotational speed will increase as well as the friction torque until we reached an equilibrium. Similarly, an increase of the airflow speed also generates an increase of the drag forces, so the rotational speed increases logically with the airflow speed. The Figure 16b shows the effects of the miniaturization of a Savonius wind turbine. As previously, the reduction of the wind turbine size leads to a reduction of the drag forces and the motor torque associated. The friction couple of the bearings is thus more difficult to offset, which implies a higher startup speed and a lower tip speed ratio. On the same Figure 16b, we can see that the carter mentioned in section 1.1 is beneficial and enables to increase slightly the rotational speed of the rotor.

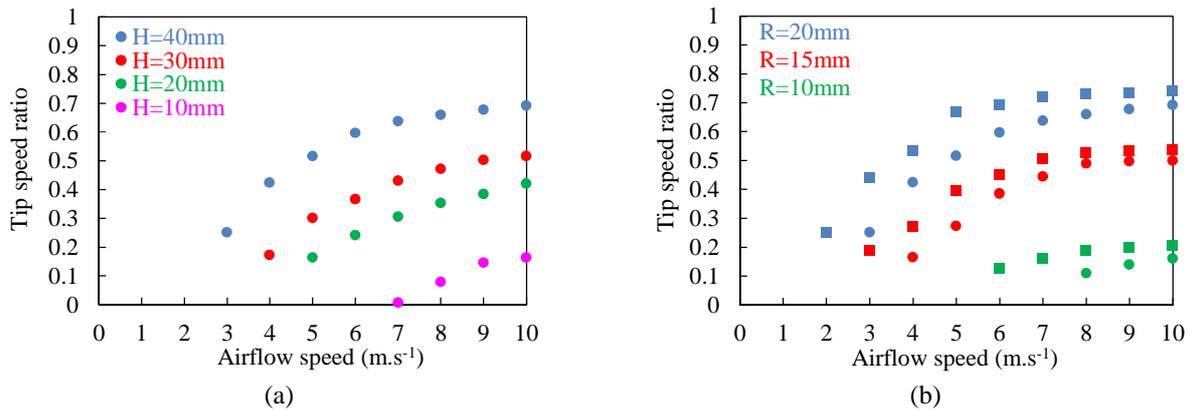

(a) (b)

**Figure 16: (a) Effect of the turbine height of a Savonius wind turbine on the tip speed ratio (R=20mm without carter). (b) Effect of the miniaturization on the tip speed ratio (H=2R). Circular points: without carter, square points: with a 2×90°carter (Figure 2a).**

We manufactured five prototypes possessing tip blade angles β(R) of 10°, 15°, 20°, 25° and 30° respectively designed for tip speed ratio equals to 2.4, 1.7, 1.3, 1.0 and 0.8. These prototypes were tested without load (friction torque of the bearing solely) between 0 and 10m.s$^{-1}$. At higher speeds (U>3m.s$^{-1}$), the prototype with β(R)=10° presents the highest tip speed ratio (1.6<λ<1.8), which augurs good power characteristics. Below 3m.s$^{-1}$, the tip speed ratio and the performances of this micro-wind turbine falls because the aerodynamic forces are not large enough to offset the frictions of the bearing. At 1m.s$^{-1}$ and 2m.s$^{-1}$, these are the prototypes with β(R)=20° and 15° which present the higher tip speed ratios with respectively λ=1.4 and 1.5 (Figure 17a).

We are also interested in the influence of the blade number which appears to be very limited as we can see in Figure 17b. Indeed, given the presence of the electrodes around the rotor, the tip losses are substantially reduced and the performances of the micro-wind turbine are consequently quasi-independent of the number of blades. For these different reasons, we will favor thereafter prototypes with β(R)=20° and $N_p$=4, in order to get good characteristics at low speed and to be able to integrate an electrostatic converter with a substantial area 2πR×H.

When we miniaturize a device by a factor k, the kinetic power received and the surface of the electrostatic converter will directly decrease by a factor k$^2$. From the manufacturing point of view, the electrets used and the voltage associated remain the same, as well as the bearings and their specific friction torque. Regarding the aerodynamics, the miniaturization of the devices is unfavorable because this leads to a diminution of the Reynolds number, hence a decrease of the aerodynamic forces creating the motor torque and an increase of the viscous drag coefficient. The startup speed will rise and the tip speed ratio will reduce (Figure 18a) which will quickly make the miniaturization very limited for ultra-low speed energy harvesting applications.

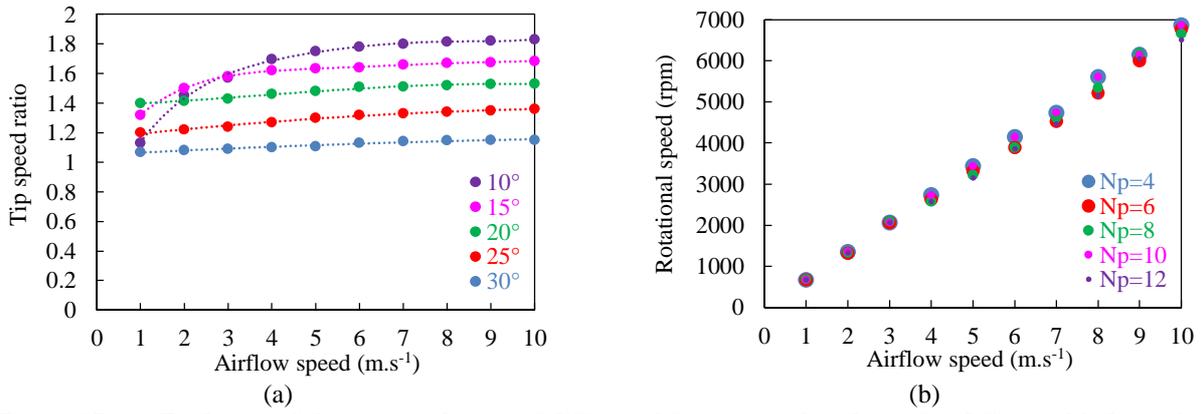

**Figure 17: (a) Evolution of the tip speed ratio of different lift-type wind turbines for different blade angles (R=20mm, $N_p$=4). (b) Impact of the number of blades on the no-load rotational speed (R=20mm, $\beta(R)$=20°, $N_p \times H/R$=2).**

The strong resisting torque due to the friction of the membranes on the stator leads to a lower rotational speed operation. It consequently implies a lower tip speed ratio that requires the use of higher blade angles. Figure 18b shows the evolution of the tip speed ratio of five prototypes ($\beta(R)$ respectively equal to 20°, 25°, 30°, 35° and 40°). These different prototypes have been tested without load ($N_m$=0), with $N_m$=1 membrane and $N_m$=2 Teflon FEP membranes; L≈πR/N=3.14cm, H=1cm and h=50µm for each membrane. First of all, we can point out the strong reduction of the tip speed ratio with the rise of the number of membranes. At 10m.s$^{-1}$ for example, the tip speed ratio is respectively equal to λ=1.5, 0.75 and 0.25 for $N_m$=0, 1 and 2. In addition, this increase in the frictional surface also tends to push up the startup speed from 1m.s$^{-1}$ to 4-5m.s$^{-1}$. At high speed, it turns out that the prototype with $\beta(R)$=20° presents the higher tip speed ratio regardless of the number of membranes. At lower speed, by contrast, lower blade angles are better suited to reach a higher rotational speed: the optimal blade angle is typically $\beta(R)$=40° at 4-5m.s$^{-1}$.

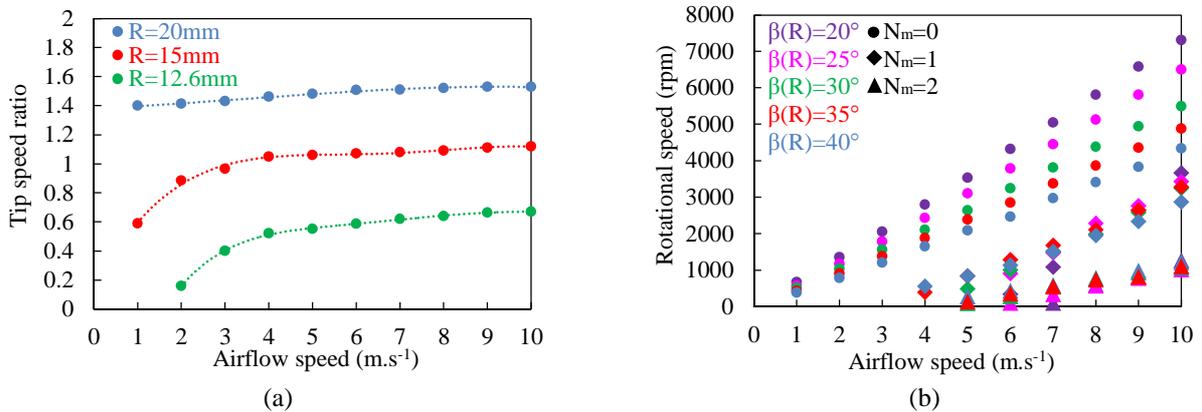

**Figure 18: (a) Effect of the miniaturization on the tip speed ratio (H/R=0.5, $\beta(R)$=20°, $N_P$=4). (b) Evolution of the no-load rotational speed of several triboelectric lift-type wind turbines for different blade angles and different number of membranes.**

### 3.5 Output power and efficiency

Figure 19a shows the best results achieved on an electret-based Savonius wind turbine. The large surface of conversion coupled to the low rotational speeds mentioned above enable us to extract from 30µW at 2m.s$^{-1}$ to 400µW at 10m.s$^{-1}$ for 125µm-thick Teflon FEP electrets previously charged at -1400V. These results have been obtained with a Savonius wind turbine of 2cm in radius, 4cm in height and a carter. The efficiency drops as the speed increases: from 0.4% at 2m.s$^{-1}$ to 0.04% at 10m.s$^{-1}$ (Figure 19a).

As regards the electret-based lift-type wind turbines, the prototype of 20mm radius presents a maximum efficiency of 9% at 1m.s$^{-1}$ (66µW). The 100µW barrier is reached at 1.5m.s$^{-1}$ as we can see in Figure 19a and 1.7mW can be produced at 10m.s$^{-1}$ ($C_p$=0.23%). The prototype of 15mm radius has also been tested and allows to extract only 4µW at 1m.s$^{-1}$, 40µW at 2m.s$^{-1}$ ($C_p$=1.25%) and a maximum power of 410µW at 10m.s$^{-1}$ (Figure 19b).

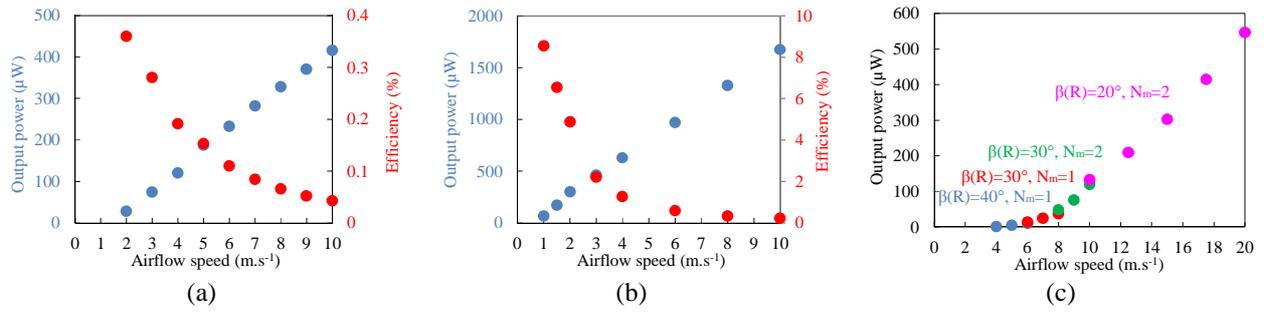

**Figure 19. Evolution of the average electrical power and the efficiency between 0 and 10m.s$^{-1}$: (a) electret-based drag-type Savonius wind turbine (R=20mm, H=40mm, N$_p$=2, N=4, V$_e$=-1400V), (b) electret-based lift-type wind turbine (R=20mm, H=10mm, N$_p$=4, N=4, e≈690µm, V$_e$≤1400V). (c) Evolution of the average electrical power of several triboelectric-based lift-type wind turbines between 0 and 20m.s$^{-1}$ (R=20mm et 2N=4 electrodes).**

Finally, the results obtained by the triboelectret-based lift-type wind turbine of 20mm radius are shown in Figure 19c. In terms of power, 100µW can be reached between 9m.s$^{-1}$ and 10m.s$^{-1}$, for a maximum power of 550µW at 20m.s$^{-1}$.

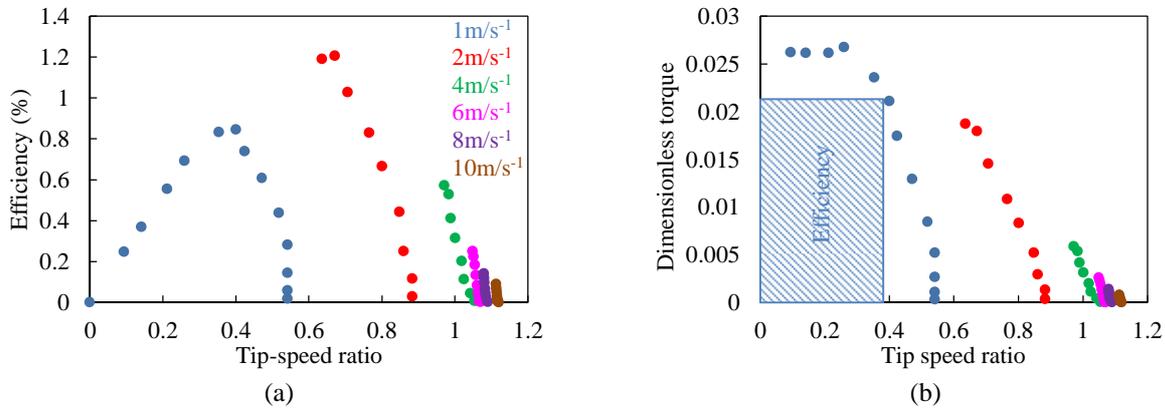

**Figure 20. Results achieved on an electret-based lift-type wind turbine (R=15mm, H=7.5mm, N$_p$=4, N=4, e≈440µm, V$_e$≤1400V). (a) Efficiency η vs tip-speed ratio λ at different speeds, (b) Dimensionless torque C$_{adim}$=η/λ vs tip-speed ratio λ at different speeds.**

### 3.6 Comparison with the state of the art

Figure 21 shows all the current results on rotating devices, including our results (round markers) and results from the state of the art (triangular markers) for airflow speeds ranging between 0 and 20m.s$^{-1}$. This figure illustrates once again the fact that all the devices from the state of the art provide overall efficiencies included between 1% and 10% for operating velocities higher than 2m.s$^{-1}$. At higher speed, our devices just allowed to extract a small fraction of kinetic energy available, i.e. between 0.01% and 1%. However, they are clearly more appropriate to ultra-low speeds with overall efficiencies between 1 and 10% from 1 to 4m.s$^{-1}$. In particular, the electret-based micro wind turbines have provided an operation from 1 to 2m.s$^{-1}$, with power flux densities ideally between 5µW.cm$^{-2}$ and 24µW.cm$^{-2}$ (red round markers).

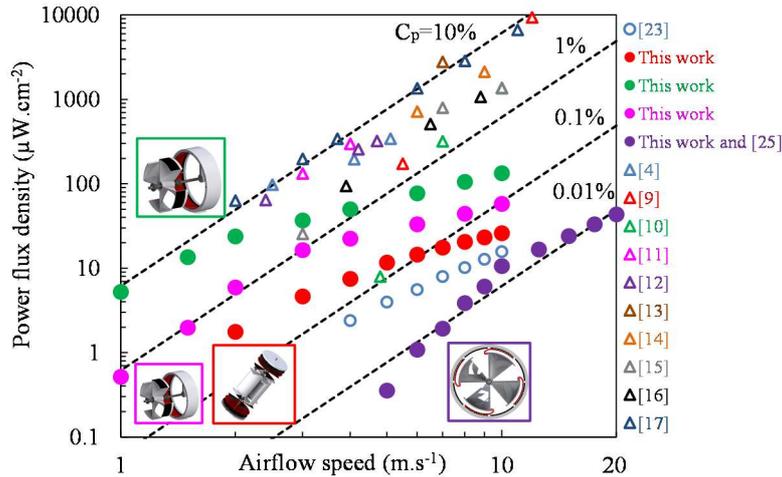

**Figure 21. Comparison with different rotational harvesters of the state of the art.**

## 4. Power management electronics

The three devices previously presented have been coupled to several power management circuits in order to supply autonomous and communicating sensors. In this last section, we propose to detail the operating principle of these different circuits and to compare them in the particular case of the triboelectric lift-type wind turbine.

### 4.1 Passive circuit

The use of a passive circuit is the simplest option and also has the merit of not consuming electrical energy (hence the term passive). In practice, it is sufficient to link the harvester to a diode bridge (to rectify the voltage), and then to a storage capacitor (Figure 22). The storage capacitor can finally be linked to a sensor (WSN in Figure 22) which can be simply controlled by the voltage across the storage capacitor.

If we want to maximize the electrical power extract by this type of circuit, the voltage across the storage capacitor must be equal to half of the harvester open-circuit voltage (at least 150V). It would therefore be necessary to use a small storage capacitor (a few dozen nanofarads) in order to quickly reach this optimal operating point. In practice, this is not conceivable because we want to supply sensors which operate at 2-3 volts. As a consequence, we have to use larger storage capacitors (a few dozen microfarads) in order to obtain a voltage across the storage capacitor ranged from a few volts to the detriment of energy efficiency.

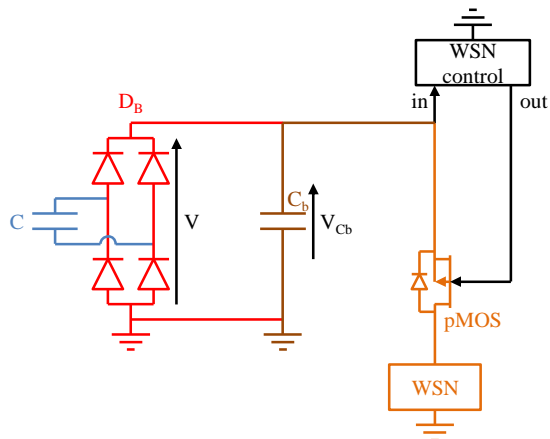

**Figure 22. Passive circuit involving the harvester in blue, the diode bridge in red, the storage capacitor in brown and the Wireless Sensor Network (WSN) in orange.**

### 4.2 MPP circuit

As we just saw, the passive strategy is very limited in our case due to the big difference of voltage between the input (the harvester) and the output (the sensor). In order to solve this problem, we can simply separate the input and the output by a transformer (purple in Figure 23).

This is how the MPP (Maximum Power Point) circuit was designed: the harvester is still linked to the diode bridge which is itself linked to a small capacitor $C_0$ of a few dozen nanofarads. This design allows to obtain a high input optimal voltage (V) and a low voltage of a few volts at the output of the transformer output ($V_{Cb}$). The "flyback control" enables to continuously transfer the energy through the transformer in order to maintain an optimal voltage across $C_0$. As indicated earlier, the supply of the sensor is controlled by the low voltage across the storage capacitor $C_b$ of a few dozen of microfarads. Initially, this circuit had been designed during a previous study because it allowed to simply parallelize the electrostatic converters. Additional informations can be found in [41].

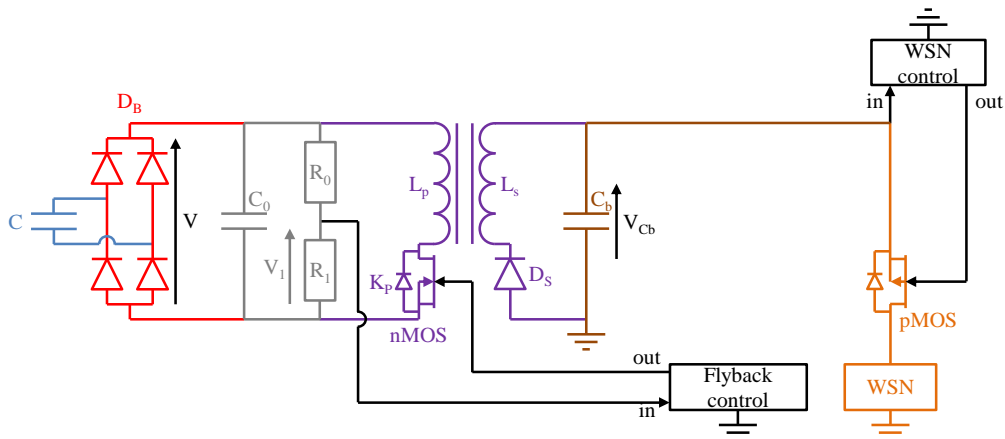

**Figure 23. High voltage MPP circuit comprising an input capacitor and a voltage divider in grey and a flyback in purple.**

### 4.3 SECE circuit

A third option consists in using an active SECE (Synchronous Electric Charge Extraction). In this case, we do not need an input capacitor before the transformer because the energy is directly and synchronously transferred through the transformer at each maximum voltage on the sinusoidal input voltage V. This technic is very commonly used in energy harvesting and theoretically allows to increase the output power by four [42] compared to the two other circuits.

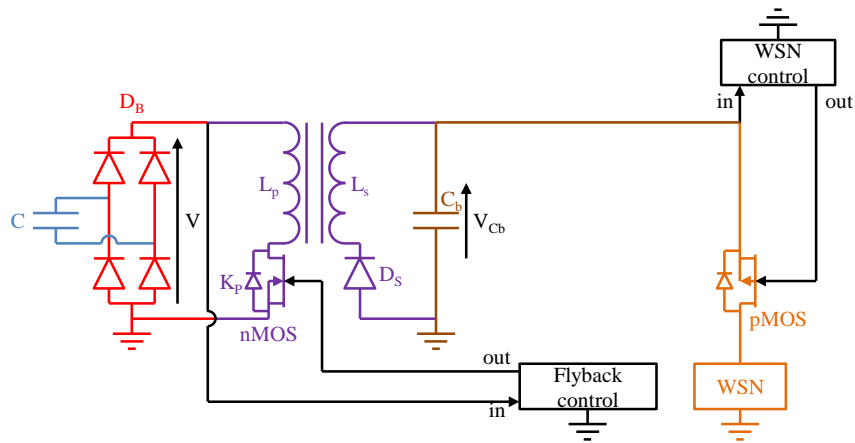

**Figure 24. SECE circuit.**

### 4.4 Supply of a temperature/acceleration/magnetic field sensor

In order to complete the energy harvesting scheme, we used a temperature sensor and an accelerometer/magnetometer linked to a low power transmission (ST's SPIRIT1 RF chip). This circuit has been designed and programmed according to a low power strategy and connected to the power management circuit in place of the "WSN". Each measurement (temperature, acceleration and magnetic field) and its wireless transmission (868MHz) operates when $V_{Cb}$ reaches $V_{Cb+}=3.2V$ (Figure 25a) and consumes around 270μJ.

This publication presents the results obtained with the lift-type triboelectric wind turbine (R=20mm, β(R)=20°, $N_p$=4, $N_m$=2) tested at 12m.s$^{-1}$ with four types of circuits: a passive circuit composed of a simple diode bridge and a storage capacitor $C_b$, the active SECE circuit, the active MPP circuit previously presented and a commercial active circuit (LTC3588), a reference for the high voltage piezoelectric energy harvesters ($V_{open}$~50V). The electronic components used during these tests are summarized in Table 2. The passive circuit allows to perform a first measurement 3 minutes and 37 seconds after the startup of the harvester (Figure 25b) while the active SECE and MPP circuits are equivalent and almost 7 times faster (respectively 30 seconds et 34 seconds). After the first measurement, the passive circuit allows to perform a measurement (270µJ) every 1 minute (Figure 25b), the SECE circuit and the MPP circuit every 6 seconds (Figure 25a). In our case, we can conclude that, despite the transition startup period and the consumption of a few microwatts (around 1.2µA under 2-3V) by the overall active circuit (flyback control + WSN control), the active energy transmission is much more interesting than a passive strategy.

| Components | Values/references |
|---|---|
| $D_b$, $D_{bp}$ | MMBD1503 |
| $C_b$, $C_s$, $C_0$ | 100µF, 10µF, 44nF |
| $R_0$, $R_1$ | 10GΩ, 100MΩ |
| Flyback of the MPP circuit | WE-Midcom: $L_p$=500µH, $L_s$=14µH |
| Flyback of the SECE circuit | Magnetic core EFD20-3F3: $L_p$=11.5mH (150 turns), $L_s$=51µH (10 turns) |

**Table 2. List of the components used.**

However, the LTC3588 active circuit proposed by Linear Technology has no interest in this case and is even outperformed by the passive circuit with a period of 7 minutes and 53 seconds to perform a first measurement (Figure 25b). Indeed, this circuit acts as our MPP circuit with a close difference that it maintains a voltage of 5V on the input capacitor. If this voltage is close to the open circuit voltage of some piezoelectric harvesters, it is far from the open circuit voltage of our electrostatic harvesters ($V_{opt}=V_{open}/2$ on the order of a few hundred volts).

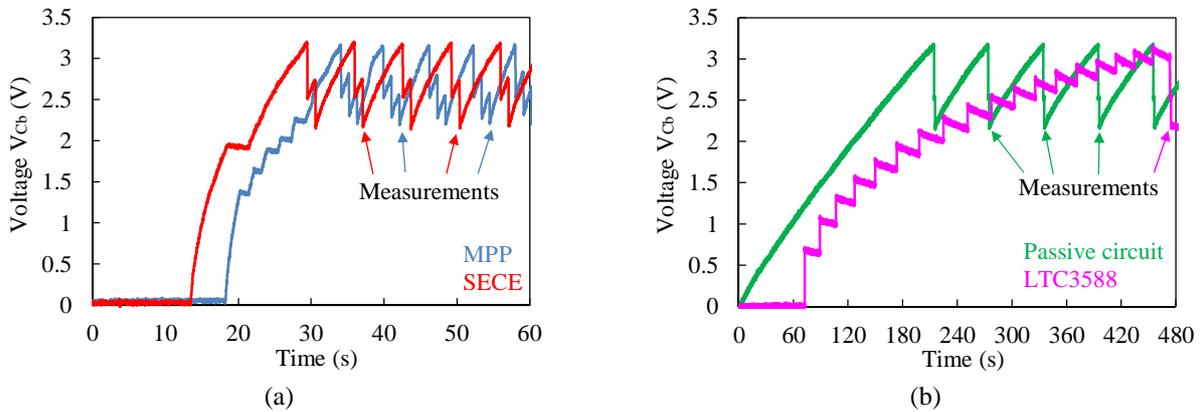

**Figure 25. Measurements of temperature, acceleration and magnetic field at 12m.s$^{-1}$ with (a) a SECE circuit and a MPP circuit, (b) a passive circuit and a commercial active circuit (LTC3588).**

## 5. Conclusions

In this paper, we laid the theoretical bases of the electrostatic micro-turbines. We have also experimentally tested several cm-scale devices whose power levels allowed us to consider different applications of autonomous sensors. Electret-based wind turbines are well suited to target very low speed applications. Among these rotational "contactless" devices, the lift-type HAWTs have shown much better performances than the drag-type VAWTs, with an operation from 1m.s$^{-1}$, higher rotational speeds (7300rpm vs 3500rpm at 10m.s$^{-1}$) as well as power levels (300µW vs 28µW at 2m.s$^{-1}$) and efficiencies (9% vs 0.4%) and not forgetting a much smaller overall volume (12.6cm$^3$ vs 50.3cm$^3$). Nevertheless, electrets are subject to their own discharge for different reasons, which is fairly problematic with our intentions to develop devices able to operate for several decades without human intervention.

The addition of membranes on the periphery of the rotor allows to solve this problem of discharge thanks to a self-polarization of the converter. However, their inclusion also causes frictions which offset the startup speed of the harvesters to 4m.s$^{-1}$ and strongly decrease the efficiencies associated. These devices still offer interesting power levels at higher speed, including the possibility to supply an autonomous sensor with RF communication (temperature/acceleration/magnetic field).

Finally, one device has been tested with different active power management circuits (MPP and SECE) which turned out to be very beneficial compared to a passive strategy and more efficient than the commercial active circuit LTC3588.